\documentclass[aps,superscriptaddress, showpacs,preprintnumbers, superscriptaddress, nofootinbibt,twocolumn]{revtex4}
\usepackage{eurosym}
\usepackage{dcolumn}
\usepackage{bm}
\usepackage{enumerate}
\usepackage{float}
\usepackage{epstopdf}
\usepackage{amsmath}
\usepackage{bm}
\usepackage{amsfonts}
\usepackage{amssymb}
\usepackage{graphicx}
\usepackage{alphalph,mathtools}
\usepackage{etoolbox}

\def\be{\begin{equation}}
\def\ee{\end{equation}}
\def\bea{\begin{eqnarray}}
\def\eea{\end{eqnarray}}

\newcommand{\R}{\ensuremath{\mathbb{R}}}

\begin{document}

\title{Finslerian geometrization of quantum mechanics in the hydrodynamical representation}
\author{Shi-Dong Liang}
\email{stslsd@mail.sysu.edu.cn}
\affiliation{School of Physics, Sun Yat-Sen University, Guangzhou 510275, People's
Republic of China,}
\affiliation{State Key Laboratory of Optoelectronic Material and Technology, and
Guangdong Province Key Laboratory of Display Material and Technology, \\
Sun Yat-Sen University, Guangzhou 510275, People's Republic of China,}
\author{Sorin V. Sabau}
\email{sorin@tokai.ac.jp}
\affiliation{School of Science, Department of Mathematics, Tokai University, Sapporo
005-8600, Japan}
\author{Tiberiu Harko}
\email{t.harko@ucl.ac.uk}
\affiliation{School of Physics, Sun Yat-Sen University, Guangzhou 510275, People's
Republic of China,}
\affiliation{Department of Physics, Babes-Bolyai University, Kogalniceanu Street,
Cluj-Napoca 400084, Romania,}
\date{\today }

\begin{abstract}
We consider a Finslerian type geometrization of the non-relativistic quantum
mechanics in its hydrodynamical (Madelung) formulation, by also taking into
account the effects of the presence of the electromagnetic fields on the
particle motion. In the Madelung representation the Schr\"{o}dinger equation
can be reformulated as the classical continuity and Euler equations of
classical fluid mechanics in the presence of a quantum potential,
representing the quantum hydrodynamical evolution equations. The equation of
particle motion can then be obtained from a Lagrangian similar to its
classical counterpart. After the reparametrization of the Lagrangian it turns out that the Finsler metric describing the geometric properties of quantum hydrodynamics is a Kropina metric. We present and discuss in detail the metric and the geodesic equations describing the geometric properties of the quantum motion in the presence of electromagnetic fields. As an application of the obtained formalism we consider the Zermelo navigation problem in a quantum hydrodynamical system, whose solution is given by a Kropina metric. The case of the Finsler geometrization of the quantum hydrodynamical motion of spinless particles in the absence of electromagnetic interactions is also considered in detail, and the Zermelo navigation problem for this case is also discussed.
\end{abstract}

\maketitle
\tableofcontents


\section{Introduction}

The extremely successful geometrization of the gravitational interaction via
Riemannian geometry, general relativity and the Einstein gravitational field
equations has raised the important question if geometrical methods could be
also successfully applied for the geometrical description of other branches
of physics. For the Newtonian mechanics and gravity a first step in this
direction was taken by Cartan \cite{Cartan} (see also \cite{Havas}), who did
show that the motion of a particle in a gravitational field, 
can be
reformulated as a geodesic equation of motion.

Hence the geometric interpretation of classical mechanics, or, more
generally, of classical physics, is a long standing subject in both
mathematical and theoretical physics, representing a very active field of
research \cite{Pet1}-\cite{Dob}. Two main \textit{metric} approaches have
been developed for a geometric description of classical mechanics. In the
first approach, the Jacobi metric approach,
in order to describe Hamiltonian mechanical dynamical systems
one introduces a metric of the form $ds^2_J=W\left(q^a\right)\delta _{ab}dq^adq^b$,
where  $W$ represents the conformal factor given by $%
W\left(q^a\right)=E-V\left(q^a\right)$ \cite{Kau}. This metric
is called the Jacobi metric. The flow associated with a time dependent
Hamiltonian  can then be reformulated as a geodesic flow in a curved,
but conformally flat, manifold \cite{Kau}.


An alternative \textit{metric} description of classical dynamics is based on
the Eisenhart metric \cite{PR,Pet5,e2,e3,e9,e10}. In this approach one considers an
ambient space with an extra dimension, $M\times \mathbf{R}^2$, with local
coordinates $(q^0,q^1,\ldots,q^i,\ldots,q^N,q^{N+1})$. On this space we can
introduce a non-degenerate pseudo-Riemannian metric, called the Eisenhart
metric, given by
$ds^2_E = g_{\mu \nu}^{(E)} dq^{\mu}dq^{\nu} =\delta_{ij} dq^i dq^j
-2V(q)(dq^0)^2 + 2 dq^0 dq^{N+1}$
where $\mu$ and $\nu$ run from $0$ to $N+1$, while $i$ and $j$ run from 1 to
$N$. Then it can be shown that the motions of a Hamiltonian dynamical system
can be obtained as the canonical projection of the geodesics of $(M\times
\mathbf{R}^2,g^{(E)}$ on the configuration space-time, $\pi : M\times
\mathbf{R}^2 \mapsto M\times \mathbf{R}$ \cite{PR}.


We may call the above interpretations of physical theories as the
{\it geometro-dynamical approach}. Such approaches are known from a long time in
the framework of the general geometric theory of the dynamical systems.

Perhaps the most important field of study of theoretical physics, quantum
mechanics, can also be formulated as a geometric theory, in which standard
methods of differential geometry play an important role.
Quantum
mechanics can be cast into a classical Hamiltonian form in terms of a
symplectic structure, not on the Hilbert space of state-vectors but on the
more physically relevant infinite-dimensional manifold of instantaneous pure
states \cite{Kibble1}. This geometrical structure can accommodate
generalizations of quantum mechanics, including nonlinear relativistic
models \cite{Kibble2}. A metric tensor from the underlying Hilbert space
structure for any submanifold of quantum states was defined in \cite{VP}.
The case where the manifold is generated by the action of a Lie group on a
fixed state vector (generalized coherent states manifold) was studied in
detail.
In \cite{An} it was argued that quantum mechanics is
fundamentally a geometric theory
Using the natural metric on the projective space the Schr\"{o}dinger's equation for an isolated system can be reformulated in geometric
terms. On the other hand the manifold of pure quantum states can be regarded
as a complex projective space endowed with the unitary-invariant Fubini-
Study metric \cite{BrHu}.

A geometrization of the non-relativistic quantum mechanics for mixed states
was introduced in \cite{GiSo}, making use of the Uhlmann's principal fibre
bundle.
A formulation of the
formalism of quantum mechanics in a geometrical form based on the K\"{a}hler
structure of the complex projective space was proposed in \cite{Mo}.
A geometric framework for mixed quantum states represented by
density matrices,
was discussed in \cite{Hey}. A geometric
setting of quantum variational principles and their extension to include the
interaction between classical and quantum degrees of freedom was considered
in \cite{BoTr}. The construction of a general prescription to set up a
well-defined and self-consistent geometric Hamiltonian formulation of
finite-dimensional quantum theories, where phase space is given by the
Hilbert projective space (as a K\"{a}hler manifold) was considered in \cite%
{Past}.

One of the important extensions of Riemannian geometry is the Finsler
geometry \cite{Fins, Rund, Asan, Bej, book2,book3}. A Finsler space is based on the general line element $ds=F\left(x^1,x^2,...,x^n;dx^1,dx^2,...,dx^n\right)$, where $F(x,y)>0$ for $y\neq 0$ is a function on the tangent bundle $T(M)$, and homogeneous of degree 1 in $y$.  In fact Finsler geometry is not just another generalization of Riemannian geometry, but it's just Riemannian geometry without the restriction that the line element is quadratic \cite{Chern}. If in Riemannian geometry we have $F^2=g_{ij}(x)dx^idx^j$, in a Finsler space the length of a differential line element at $x$ depends in general on both $x$ and $y$ according to $\sqrt{d\vec{x}\cdot d\vec{x}}=\left[g_{AB}\left(x,y\right)dx^Adx^B\right]^{1/2}$. Finsler geometry has many important applications in electromagnetism,  gravitation or continuum mechanics \cite{Ikeda, Brandt, Suh, Clay}.
 A Finslerian approach to classical mechanics was
developed in \cite{Mi1, Mi2,MHSS, MiFr05}, respectively.

One of the classic problems in optimization theory is the Zermelo navigation problem \cite{Zerm}, requiring the extremization of the travel time of an aeroplane in the presence of a wind. As shown in \cite{Rob}, the solution of the Zermelo  problem is equivalent to the determination of the minimal time travel curves in a Finsler type geometry with a Randers metric \cite{Rand}. More recently it was shown that Kropina metrics are actually singular solutions of the Zermelo navigation problem  \cite{YS1}. The Zermelo navigation problem and its applications have been intensively investigated in both the mathematical and physical literature \cite{Zerm1, Zerm2,Zerm3,Zerm4,Zerm5}. The Zermelo navigation problem has been also considered in the fields of quantum mechanics and quantum computation \cite{Zq1, Zq2,Zq3,Zq4,Zq5}.  The quantum navigation problem can be formulated as finding the time-optimal control Hamiltonian, which transports a given initial state to a target state in the presence of a quantum wind, that is, under the influence of external fields or potentials \cite{Zq4}. It is possible to obtain a universal quantum speed limit by lifting the problem from the state space to the space of unitary gates. The optimal times for implementing unitary quantum gates in a constrained finite dimensional controlled quantum system were analyzed in \cite{Zq2}. A Randers metric,
was constructed, the geodesics of which are the time optimal trajectories compatible with the prescribed constraint.

Recently, a Finslerian type geometrization of quantum mechanics
that describes the time evolution of particles as geodesic lines in a curved
space was proposed in \cite{Tav1}, in which the curvature of the Finsler space is induced by the quantum potential.
A description of
the phenomenon of self-interference using Finslerian geometrical formulation
was discussed in \cite{Tav2}. This formalism removes the need for the
concept of wave function collapse in the interpretation of the act of
measurement, that is, of the emergence of the classical world.
The Finslerian geometrical formulation of quantum field theory in space-time was unified with classical Einstein's general relativity in \cite{Tav2a}. 
In \cite{Tav3} it was shown that for a system of two entangled particles, there is a
dual description to the particle equations in terms of classical theory of
conformally stretched spacetime. These entangled particle equations can be
interpreted via the framework of Finsler geometry.
A geometrization of quantum hydrodynamics was discussed in \cite{Foskett}.

It is the goal of the present paper to consider a Finsler type
geometrization of quantum mechanics along the lines suggested in \cite%
{Tav1,Tav2} for {\it a charged particle under the influence of an exterior
electromagnetic field}. After reformulating, with the use of the Madelung
representation of the wave function \cite{Mad, Take}, the Schr\"{o}dinger equation
as a quantum hydrodynamical system, it follows that the equation of motion
of the quantum particles can be derived from a classical action principle,
with the Lagrangian containing the kinetic energy of the particle, the electromagnetic potentials and exterior potentials, as well as the quantum potential $V_Q$, which determines the quantum properties of the dynamical evolution.

With the help of a
particular reparametrization of the time coordinate one can associate to
this Lagrangian a {\it Finsler type fundamental function, and a Finsler type fundamental
metric}. It turns out that the corresponding Finsler fundamental function belongs to the class of $(\alpha, \beta)$  metrics \cite{Mat1,Mat2}, or, more exactly, to the Kropina metrics \cite{Krop1, Krop2}. Hence {\it the Finsler type representation of quantum mechanics can be discussed in the framework of a well known geometric approach}. The Finslerian approach allows us to reformulate the equation of motion of quantum
particles as a geodesic equation. {\it Moreover, the important Zermelo quantum navigation problem has an immediate solution within the quantum hydrodynamical approach, and the Finslerian representation of quantum mechanics}. We discuss in detail the Finsler geometric approach to quantum hydrodynamics for both charged particles in the presence of electromagnetic interactions, as well as the limiting case of neutral spinless particles. In all cases the corresponding Finsler metrics and geodesic equations are obtained.

The present paper is organized as follows. In Section~\ref{sect2} we
introduce the hydrodynamic formulation of quantum mechanics for charged
particles in the presence of electromagnetic interactions. In Section~\ref%
{sect3} we briefly review the basic mathematical properties and definitions
of the Finsler geometries, of the $\left(\alpha, \beta\right)$ and Kropina
metrics, and of the Zermelo navigation problem, respectively. The Finslerian
quantization of the hydrodynamic version of quantum mechanics in the
presence of electromagnetic fields in presented in Section~\ref{sect4}. The solution of the
Zermelo navigation problem  in the hydrodynamical formulation of quantum
mechanics is considered in Section~\ref{sect5}. The case of the Finslerian
geometrization of the standard hydrodynamic representation of the Schr\"{o}%
dinger for the case of the neutral particle is presented in detail in
Section~\ref{sect6}. Finally, we discuss and conclude our results in Section~%
\ref{sect7}.

\section{Hydrodynamical formulation of non-relativistic quantum mechanics}\label{sect2}

In the presence of an external electromagnetic field the quantum dynamics of
a spinless particle with mass $m$ is described by the Schr\"{o}dinger
equation given by
\begin{eqnarray}
i\hbar \frac{\partial \Psi \left( \vec{r},t\right) }{\partial t} &=&\frac{1}{%
2m}\left[ \frac{\hbar }{i}\nabla -\frac{e}{c}\vec{A}\left( \vec{r},t\right) %
\right] ^{2}\Psi \left( \vec{r},t\right) +  \notag  \label{Sch} \\
&&e\phi \left( \vec{r},t\right) \Psi \left( \vec{r},t\right) +V\left( \vec{r}%
,t\right) \Psi \left( \vec{r},t\right) ,
\end{eqnarray}%
where $\phi \left( \vec{r},t\right) $ and $\vec{A}\left( \vec{r},t\right) $
are the scalar and vector potential of the electromagnetic field, and $%
V=V\left( \vec{r},t\right) $ is the potential of the non-electromagnetic
forces. The electric and magnetic fields can be obtained from the
electromagnetic potentials as $\vec{B}=\nabla \times \vec{A}$ and $\vec{E}%
=-e\nabla \phi -\dot{\vec{A}}/c$, respectively \cite{LL}. From the Schr\"{o}%
dinger equation one can obtain immediately the conservation law for the
probability current density as given by
\begin{equation}
\frac{\partial \left\vert \Psi \left( \vec{r},t\right) \right\vert ^{2}}{%
\partial t}+\nabla \cdot \vec{j}=0,
\end{equation}%
where
$
\vec{j}=\left( \Psi ^{\ast }\hat{\vec{v}}\Psi +\Psi \hat{\vec{v}}%
^{\ast }\Psi ^{\ast }\right) /2=\mathrm{Re}\left( \Psi ^{\ast }\hat{\vec{v}}%
\Psi \right)$, and
$\hat{\vec{v}}=\left( \hat{\vec{p}}-e\vec{A}/c\right)/m$.

In general the operators $\hat{\vec{p}}=\left( \hbar /i\right) \nabla $ and $%
\vec{A}$ do not commute, so that $\left[ \hat{\vec{p}},\vec{A}\right] =-i\hbar
\partial _{i}A_{i}=-i\hbar \nabla \cdot \vec{A}$. In order to make these
operators commutative in the following we will adopt for the vector
potential of the magnetic field the Coulomb gauge $\nabla \cdot \vec{A}=0$.

As a next step we introduce the Madelung (quantum hydrodynamical)
representation of the wave function as
\begin{equation}  \label{Mfunc}
\Psi \left(\vec{r},t\right)=R\left(\vec{r},t\right)e^{iS\left(\vec{r}%
,t\right)/\hbar},
\end{equation}
where both $R$ and $S$ are real quantities. For a given wave function $\Psi$%
, this representation does not determine $\Psi $ uniquely, since $S^{\prime
}=S+nh/2$ and $R^{\prime }=\left(-1\right)^nR$, where $n\in \mathrm{N}$
gives the same $\Psi$. However, if we restrict our analysis to non-negative $%
R$, then $R$ is uniquely determined \cite{Take}. Moreover, $S$ is also
undetermined at the nodal point $R=0$. By substituting the wave function as
given by Eq.~(\ref{Mfunc}) into the Schr\"{o}dinger Eq.~(\ref{Sch}), after
separating the real and imaginary parts we obtain \cite{Take, Holland, Pet, Pit, Khal}
\begin{equation}  \label{h1}
R\left[\frac{\partial S}{\partial t}+\frac{1}{2m}\left(\nabla S-\frac{e}{c}%
\vec{A}\right)^2+e\phi+V\right]-\frac{\hbar ^2}{2m}\Delta R=0,
\end{equation}
and
\begin{equation}  \label{h2}
\frac{\partial R^2}{\partial t}+\nabla \cdot\left[\frac{R^2}{m}\left(\nabla
S-\frac{e}{c}\vec{A}\right)\right]=0,
\end{equation}
respectively. 
By
introducing the velocity $\vec{v}$ of the quantum particle, defined as
$m\vec{v}=\nabla S-e\vec{A}/c$,
and by denoting $\rho \left(\vec{r},t\right) =R^2\left(\vec{r},t\right)=\Psi ^*\left(\vec{r}%
,t\right)\Psi \left(\vec{r},t\right)$,
Eqs.~(\ref{h1}) and (\ref{h2}) take the form
\begin{equation}  \label{h3}
\frac{\partial S}{\partial t}+\frac{1}{2}m\vec{v}^2+e\phi +V+V_Q=0,
\end{equation}
\begin{equation}
\frac{\partial \rho}{\partial t}+\nabla \cdot \left(\rho \vec{v}\right)=0,
\end{equation}
where we have introduced the quantum potential $V_Q$ defined as
\begin{equation}
V_Q=-\frac{\hbar ^2}{2m}\frac{\Delta R}{R}=-\frac{\hbar ^2}{2m}\left[\frac{%
\Delta \rho}{\rho }-\frac{\left(\nabla \rho\right)^2}{2\rho}\right].
\end{equation}
By taking the gradient of Eq.~(\ref{h3}) we obtain the equation of motion of
quantum particle as
\begin{equation}  \label{h4}
m\frac{d\vec{v}}{dt}=\vec{F}-\nabla \left(V+V_Q\right),
\end{equation}
where $d/dt=\frac{\partial }{\partial t}+\vec{v}\cdot \nabla $ and $\vec{F}=e%
\vec{E}+\left(e/c\right)\vec{v}\times \vec{B}$ is the Lorentz force acting
on the particle. Eq.~(\ref{h4}) can be rewritten as
\begin{equation}
\frac{dv_i}{dt}=\frac{1}{m}F_i-\frac{1}{m\rho}\sum _{k}\frac{\partial \sigma
_{ik}}{\partial x_k}-\frac{1}{m}\frac{\partial V}{\partial x_i},
\end{equation}
where $\sigma _{ik}=-\frac{\hbar ^2}{4m}\frac{\partial ^2\ln \rho}{\partial
x^i\partial x^k}$.

We introduce now the quantum trajectories via the definition
$\vec{v}=d\vec{r}/dt$ \cite{Holland, Bohm1, Bohm2}, which allows us to write the equation of motion (\ref{h4}) in the form
\begin{equation}  \label{h5}
m\frac{d^2\vec{r}}{dt^2}=\vec{F}-\nabla \left(V+V_Q\right).
\end{equation}


\section{A brief review of Finsler geometry, general $(\alpha,%
\beta)$-metrics, and of Zermelo navigation}\label{sect3}

In the present Section we will briefly review the basics of the Finsler
geometry, of the $(\alpha,\beta)$-metrics, and of the Zermelo navigation
problem.

Finsler geometry naturally appears in the framework of classical mechanics when dissipative effects are taken into account. Let's assume that the
equations of motion of a dynamical system, defined on an $n$-dimensional differentiable manifold $M$, can be obtained from a Lagrangian $%
L$ by using the Euler-Lagrange equations, given by
\begin{equation}
\frac{d}{dt}\frac{\partial L}{\partial y^{i}}-\frac{\partial L}{\partial
x^{i}}=F_{i},i\in \left\{1,2,...,n\right\},  \label{EL}
\end{equation}
where $F_{i}$ are the external forces \cite{MHSS}. We call
the triplet $\left( M,L,F_{i}\right) $ a Finslerian mechanical system \cite%
{MiFr05}. If the Lagrangian $L$ is regular, then the above Euler-Lagrange
equations are equivalent to a system of second-order differential equations
of the form
\begin{equation}
\frac{d^{2}x^{i}}{dt^{2}}+2G^{i}\left( x^{j},y^{j},t\right) =0,i\in \left\{
1,2,...,n\right\} ,  \label{EM}
\end{equation}
where in a neighborhood of some initial conditions $\left( \left( x\right)
_{0},\left( y\right) _{0},t_{0}\right) $ each function $G^{i}\left(
x^{j},y^{j},t\right) $ is $C^{\infty }$ in $\Omega $. These equations can be naturally interpreted as describing geodesic motion in a Finsler space.

\subsection{Finsler geometry}

One of the fundamental assumptions of modern theoretical physics is that
spacetime can be described mathematically as a four dimensional
differentiable manifold $M$, endowed with a pseudo-Riemannian tensor $g_{I
J} $, where $I,J,K...=0,1,2,3$. Then the space-time interval between two
events $x^{I}$ and $x^{I} + dx^{I}$ on the world line of a standard clock is
defined by the chronological hypothesis to be $ds=\left(g_{I
J}dx^{I}dx^{J}\right)^{1/2}$ (using Einstein's summation convention) \cite%
{Tava,Tava1}. One of the most important metrical generalizations of the Riemannian
geometry that was already hinted by Riemann himself \cite{Riem} is the
geometry subsequently initiated by Finsler \cite{Fins} (for detailed description of
Finsler geometry see \cite{Rund, Asan, Bej, book2,book3}).

 Finsler spaces are metric
spaces in which the interval $ds$ between two infinitesimally close points $%
x=(x^{I})$ and $x+dx=(x^{I} + dx^{I})$ is given by
\begin{equation}  \label{dsF}
ds=F\left(x,dx\right),
\end{equation}
where the Finsler metric function $F$ is positively homogeneous of degree
one in $dx$, so that
\begin{equation}
F\left(x,\lambda dx\right)=\lambda F\left(x,dx\right).
\end{equation}
In order to allow the use of local coordinates in computations, the Finsler
metric function $F$ can be written in terms of $(x,y)=(x^I,y^I)$, the
canonical coordinates of the tangent bundle, where $y=y^I\dfrac{\partial}{%
\partial x^I}$, is any  tangent vector $y$ at $x$. The Finsler tensor $%
g_{I J}$ is defined then as
\begin{equation}  \label{Hessian mat}
g_{I J}\left(x,y\right)=\frac{1}{2}\frac{\partial ^2F^2\left(x,y\right)}{%
\partial y^{I}\partial y^{J}},
\end{equation}
which allows to write Eq.~(\ref{dsF}) as $ds^2=g_{I J}\left(x,y\right)y^{I}y^{J}$.
Riemann spaces are special cases of
Finsler spaces, corresponding to $g_{IJ}\left(x,y\right)=g_{IJ
}\left(x\right)$ and $y^{I}=dx^{I}$, respectively.

With the use of the Finsler metric we can obtain the geodesic equations in a
Finsler space in the form \cite{Tava,Tava1}
\begin{equation}
\frac{d^2x^{I}}{d\lambda ^2}+2G^{I}\left(x,\dot{x}\right)=0,
\end{equation}
or, equivalently,
\begin{equation}
\frac{d^2x^{I}}{d\lambda ^2}+\Gamma^{I}_{J K}\left(x,\dot{x}\right)\frac{%
dx^{J}}{d\lambda}\frac{dx^{K}}{d\lambda}=0,
\end{equation}
where
\begin{equation}
G^{I}\left(x,\dot{x}\right)=\frac{1}{2}\Gamma ^{I}_{ JK}y^{J}y^{K}=\frac{1}{4%
}g^{IJ}\left(\frac{\partial ^2F^2}{\partial x^K \partial y^J}y^K-\frac{%
\partial F^2}{\partial x^J}\right),
\end{equation}
and $\Gamma ^{I}_{J K}$ are the analogues of the Christoffel symbols of the
Riemann geometry, defined as
\begin{eqnarray}\label{35}
\hspace{-0.3cm}\Gamma ^{I}_{J K}\left(x, \dot{x}\right)&=&\frac{1}{2}g^{I
L}\left(x,\dot{x}\right)\Bigg[\frac{\partial g_{L J}\left(x,\dot{x}\right)}{%
\partial x^{K}}  \notag \\
\hspace{-0.3cm}&&+\frac{\partial g_{L K}\left(x,\dot{x}\right)}{\partial
x^{J}}-\frac{\partial g_{J K}\left(x,\dot{x}\right)}{\partial x^{L}}\Bigg].
\end{eqnarray}

A special class of Finsler spaces, called Berwald spaces, can be obtained
when $\Gamma ^{I}_{JK }\left(x^{A},\dot{x}^{A}\right)=\Gamma^{I}_{J
K}\left(x^{A}\right) $ \cite{Rund}.

The functions $G^I$ defined above are called the \textit{coefficients of the
spray} of $F$. They are used to define the vector field $S$ on $%
TM\setminus{0}$ by
\begin{equation*}
S=y^I\frac{\partial}{\partial x^I}-2G^I\frac{\partial}{\partial y^I}.
\end{equation*}
$S$ is called the \textit{spray} induced by $F$. A curve $%
\gamma$ on $M$ is a geodesic of $F$ if and only if its canonical lift $\hat{%
\gamma}(t)=(\gamma(t),\dot{\gamma}(t))$ to $TM$ is an integral curve of $S$.

\subsection{Kropina and general $(\alpha , \beta )$ metrics}

A special kind of Finsler space is the Randers space \cite{Rand}, having
\begin{equation}
F=\left[ a_{I J }(x)dx^{I }dx^{J }\right] ^{1/2}+b_{I }(x)dx^{I },
\end{equation}%
where $a_{I J }$ is the fundamental tensor (metric) of a Riemannian space,
and $b_{I}(x)dx^{I }$ is a linear $1$-form on the tangent bundle $TM$. Later
on, Kropina \cite{Krop1,Krop2} considered Finsler spaces equipped with
metrics of the form
\begin{equation}
F\left( x,y\right) =\frac{a_{IJ }(x)y^{I }y^{J }}{b_{I }(x)y^{I }}.
\end{equation}

Generalizing these results Matsumoto \cite{Mat1,Mat2} introduced the notion
of the $(\alpha ,\beta )$ metrics as follows: a Finsler metric function $%
F(x,y)$ is called an $(\alpha ,\beta )$ metric when $F$ is a positively
homogeneous function $F(\alpha ,\beta )$ of first degree in two variables $%
\alpha \left( x,y\right) =\left[ a_{I J }(x)dx^{I }dx^{J }\right] ^{1/2}$
and $\beta \left( x,y\right) =b_{I }(x)y^{I }$, respectively. In the
following we will suppose that $\alpha $ is a Riemannian metric, that is, it
is non-degenerate (regular), and positive-definite. In the case of the
Randers and Kropina metrics we have $F=\alpha +\beta $, and $F=\dfrac{\alpha^2}{\beta}$, respectively. Hence the Randers and Kropina metrics
belongs to the class of the $(\alpha ,\beta )$ metrics. Alternatively, we
can consider general $(\alpha ,\beta )$ metrics as metrics of the form $F(\alpha
,\beta )=\alpha \phi \left( \beta /\alpha \right) =\alpha \phi \left(
s\right) $, where $\ \ s=\beta /\alpha $, and $\phi =\phi (s)$ is a $%
C^{\infty }$ positive function on an open interval $(-b_{o},b_{o})$.

In the following we introduce the quantities
$r_{I J }:=\left(b_{I |J }+b_{J |I }\right)/2$, $s_{I J }:=
\left(b_{I |J }-b_{J |I }\right)/2$,
$s_{\ J }^{I }:=a^{I K }s_{K J }$,  $s_{I }:=b_{J }s_{\ I }^{J }=b^{K
}s_{K I }$, and $e_{I J }:=r_{I J }+b_{I }s_{J }+b_{J }s_{I }$, respectively,
where $``|"$ denotes \textit{the covariant derivative with respect to the
Levi-Civita connection of $\alpha $}. In Finsler geometry it customary to
denote $r_{00}:=r_{I J }y^{I }y^{J },s_{0}:=s_{I }y^{I }$, etc.

\subsubsection{Geodesic equations for the Kropina metric}


The fundamental tensor of a Kropina space with $F=%
\dfrac{\alpha ^{2}}{\beta }=\alpha \phi (s)$, where $\phi (s)=\dfrac{1}{s}$, $s=\dfrac{\beta}{\alpha}$, reads  \cite{Shib}
\begin{equation}\label{Kropina gij}
g_{I J }=\frac{2\alpha ^{2}}{\beta ^{2}}a_{I J }+\frac{3\alpha ^{4}}{\beta
^{4}}b_{I }b_{J }-\frac{4\alpha ^{3}}{\beta ^{3}}(b_{I }\alpha _{J }+b_{I
}\alpha _{J })+\frac{4\alpha ^{2}}{\beta ^{2}}\alpha _{I }\alpha _{J}.
\end{equation}

For the geodesic spray of a general Kropina metric we obtain
\begin{equation}
\begin{split}
G^{I } =&\bar{G}^{I}-\frac{\alpha ^{2}}{2\beta }s_{\ 0}^{I }-\frac{\beta }{%
b^{2}\alpha ^{2}}\Big \{\frac{\alpha ^{2}}{\beta }s_{0}+r_{00}\Big \}{y^{I }}
\\
&+\frac{1}{2b^{2}}\Big \{\frac{\alpha ^{2}}{\beta }s_{0}+r_{00}\Big \}b^{I },
\end{split}
\label{GJH_K}
\end{equation}
where the index 0 means contraction by $y^I$.


Then the Kropina space unit speed geodesics' equations are given by
\begin{equation}
\frac{d^{2}x^{I }}{dt^{2}}+2G^{I }\left( x(t),\frac{dx(t)}{dt}\right) =0.
\end{equation}

\subsection{The Zermelo navigation problem}


We start by recalling the famous \textit{Zermelo' s navigation problem} \cite%
{Zerm}. Imagine a ship sailing on the sea in the presence of a wind of speed
$W^i$ relative to the ground. The main task of the ship's captain is to
minimize the travel time, by assuming that the ship sails at constant speed
relative to the sea. If the ship travels on a long distance the captain must
take into account the curvature of the Earth. In the absence of the wind the
optimal travel route would be a circle, that is the geodesic of the
Riemannian manifold with metric $h_{ij}$. In the following we describe the
sea by a Riemannian space $(M, h)$. {\it The pair $\left(h_{ij}, W^i\right)$ is
referred to as the Zermelo navigation data}. It was shown in \cite{Shen} that the
Zermelo problem is equivalent to finding the minimal travel time curve in a
manifold equipped with a Randers-Finsler metric with original data $%
\left(a_{ij} , b_i\right)$. The Randers geodesics are given by
\begin{equation}
\frac{d^2x^I}{ds^2}+\Gamma _{JK}^I\frac{dx^J}{ds}\frac{dx^K}{ds}=a^{IJ}F_{JK}%
\frac{dx^K}{ds},
\end{equation}
where $ds$ is the arc length of the Riemannian metric $a_{IJ}$, $\Gamma
_{JK}^I$ are the Christoffel symbols constructed from $a_{IJ}$, and $%
F_{JK}=\partial _Jb_K-\partial _Kb_J$ (\cite{book2}).

Suppose now that the ship can sail with a constant speed for a unit time on the
calm sea. Let the speed be $h$-unit speed. We denote the velocity of the
ship on the calm sea by a unit vector $u$. And let us suppose that the wind
is blowing with the $h$-unit speed, which is the same as that of the ship on
the calm sea. We denote the wind and the velocity of the ship on the windy
sea by the $h$-unit vector field $W$ and the vector $v$, respectively.
Then we have the equation $u+W=v$. From it we find $|v-W|=1$.

The length of $v$ is the speed of the ship on the windy sea for a unit time.
The above equation means that the tip of the velocity of the ship on the
windy sea lies on the $W$-translate of the indicatrix of the Riemannian space $%
(M, h)$. In other words, the indicatrix of the space in which we consider
the velocity of the ship on the windy sea is the $W$-translate of the unit
sphere of $(M, h)$.

\subsubsection{Zermelo navigation and Kropina metrics}

In this Section, we characterize a Kropina metric $F(x, y)=\alpha^2/\beta$
on $M$, where $\alpha^2=a_{IJ}(x)y^Iy^J$ and $\beta=b_I(x)y^I$, by a
Riemannian metric $h$ and a unit vector field $W$ on $M$. Since we suppose
that the matrix $(a_{IJ}(x))$ is positive definite, it follows that the
matrix $(g_{IJ}(x, y))$ is also positive definite \cite{YS1}. 


The Kropina metrics can be described as another solution of the Zermelo's
navigation problem already explained above \cite{YS1}.
For a Kropina metric $F=\alpha^2/\beta$ we define a Riemannian metric $h$
and a vector field $W$ by
\begin{equation}  \label{equation 2.2}
h_{IJ}=e^{\kappa(x)}a_{IJ},\quad W_I=\frac{1}{2}e^{\kappa(x)}b_I,
\end{equation}
where $W_I=h_{IJ}W^J$, where $\kappa(x)$ is a function of $(x^I)$ alone,
and satisfies the equation,
\begin{equation}  \label{equation 2.3}
e^{\kappa(x)}b^2=4.
\end{equation}
i.e., $|W|=1$. Then we obtain
\begin{equation}  \label{equation 2.4}
\bigg| \frac{y}{F(x, y)}-W\bigg| =1,
\end{equation}
and hence it follows that $F$ is a solution of the Zermelo's navigation problem. Observe that $h^{IJ}=e^{-\kappa(x)}a^{IJ}$,and $ W^I=b^I/2$, respectively.

Conversely, consider the metric $F(x, y)$ defined by Eq.~({\ref%
{equation 2.4}), where $|W|=1$. Solving for $F$ we obtain
\begin{equation*}
F(x, y)=\frac{|y|^2}{2h(y,W)}.
\end{equation*}
Defining $a_{IJ}$ and $b_I$ by $a_{IJ}:=e^{-\kappa(x)}h_{IJ}$ and $%
b_I:=2e^{-\kappa(x)}W_I$, respectively, we have $F(x, y)=\alpha^2/\beta$. }

Summarizing the above discussion, we obtain the following results \cite{YS1}, \cite{YS2}, \cite{SSY}.

For a Kropina space $(M, F=\alpha^2/\beta)$, where $\alpha=\sqrt{a(y,y)}$
and $\beta=b_I(x)y^I$, we define a new Riemannian metric $h=\sqrt{h(y,y)} $
and a unit vector field $W=W^I(\partial/\partial x^I)$ by Eqs.~(\ref{equation 2.2}%
) and (\ref{equation 2.3}). Then, the Kropina metric $F$ satisfies
Eq.~(\ref{equation 2.4}), and hence it is a solution of Zermelo's
navigation problem.

Conversely, suppose that $h=\sqrt{h(y,y)}$ is a Riemannian metric and $%
W=W^I(\partial/\partial x^I)$ is a unit vector field on $(M,h)$. Consider
the metric $F$ defined by Eq.~(\ref{equation 2.4}). Let $\kappa(x)$ be a
function of $(x^I)$ alone, which satisfies (\ref{equation 2.3}), and define $%
a_{IJ}$ and $b_I$ by $a_{IJ}(x):=e^{-\kappa(x)}h_{IJ}(x)$ and $%
b_I(x):=2e^{-\kappa(x)}W_I$, respectively. Then, we have $F=\alpha^2/\beta$.


\section{Finslerian geometrization of quantum hydrodynamics}\label{sect4}

In the present Section, starting from the equivalent Lagrangian formulation
of quantum hydrodynamics, we will consider a
geometric formulation of the Schr\"{o}dinger equation in the presence of the
electromagnetic interactions. In particular we will show that {\it quantum
mechanics can be interpreted as describing geodesic motion in a Kropina
space}. A similar approach for the Finsler geometrization of classical
mechanics was introduced in \cite{Shib}.

\subsection{Kropina space representation of quantum hydrodynamics}

As a fist step in our approach we consider a general Lagrangian of the form $%
L\left( t,x^{i},\dot{x}^{i}\right) $, $i=1,2,3$, to which we associate the
arc-length element $ds$ given by
\begin{equation}
ds=L\left( t,x^{i},\dot{x}^{i}\right) dt.  \label{ds}
\end{equation}%
We introduce now a new evolution parameter $\tau =\tau (t)$, which allows us
to rewrite the $ds$ as
\begin{equation}
ds=L\left( t,x^{i},\frac{\frac{dx^{i}}{d\tau }}{\frac{dt}{d\tau }}\right)
\frac{dt}{d\tau }d\tau .  \label{ds1}
\end{equation}

We introduce now the new coordinate $x^{0}=t$, and we relabel the
coordinates as $x^{I }=\left( x^{0},x^{i}\right) $, $i=1,2,3$. Moreover, we
denote $y^{I }=dx^{I }/d\tau =\left( dt/d\tau ,dx^{i}/d\tau \right) $. Then (%
\ref{ds1}) can be written as
\begin{equation}
ds=F\left( x^{I },y^{I }\right) d\tau ,  \label{ds2}
\end{equation}%
where
\begin{equation}
F\left( x^{I },y^{I }\right) =L\left( t,x^{i},\frac{y^{i}}{y^{0}}\right)
y^{0}.
\end{equation}

Eq.~(\ref{h5}), giving the equation of motion of a quantum particle in the hydrodynamic representation can be derived from  the classical Lagrangian 
\begin{equation}\label{Lag}
L\left( t,x^{i},\dot{x}^{i}\right) =\frac{1}{2}m_{ij}(x)\dot{x}^{i}\dot{x}%
^{j}+\frac{e}{c}A_{i}\dot{x}^{i}-\left( e\phi +V+V_{Q}\right),
\end{equation}%
where for the sake of generality we assume that the components of the mass matrix $m_{ij}$ are functions of $x^i$, and that the matrix $%
m_{ij}$ is positive definite. Eq.~(\ref{Lag}) will be the starting point for
our Finsler geometrization of the hydrodynamic version of quantum mechanics.

Therefore, we obtain for $F$ the expression 
\begin{eqnarray}  \label{F2}
F\left( x^{I },y^{I }\right) &=&\frac{1}{2}m_{ij}\frac{y^{i}y^{j}}{y^{0}}+%
\frac{e}{c}A_{i}y^{i}-  \notag \\
&&\left( e\phi +V+V_{Q}\right) y^{0}, i,j=1,2,3.
\end{eqnarray}

Eq.~\eqref{F2} can be rewritten as follows
\begin{eqnarray}  \label{F3}
F\left( x^{I },y^{J }\right) &=&\frac{\frac{1}{2}m_{ij}{y^{i}y^{j}} +\frac{e%
}{c}A_{i}y^0y^{i}- \left( e\phi +V+V_{Q}\right) (y^{0})^2}{ y^{0}}  \notag \\
&=&\frac{\alpha^2}{\beta}\left( x^{I },y^{J }\right) ,
\end{eqnarray}
where we have introduced the notations
\begin{equation}  \label{56}
\alpha^2:=a_{IJ}(x)y^I y^J , \quad \beta:=y^0,
\end{equation}
and $(a_{IJ})_{I,J=0,\dots,3}$ is the symmetric matrix 
with the entries
\begin{eqnarray}  \label{associated Riemann}
\hspace{-0.5cm}a_{00}(x^0,x^i)& :=&- \left( e\phi +V+V_{Q}\right), \nonumber\\
\hspace{-0.5cm}a_{0i}(x^i)& :=&\frac{e}{2c}A_{i},a_{ij}(x^i)=\frac{1}{2}m_{ij}, i\in \{1,2,3\}.
\end{eqnarray}

We will call this Riemannian metric the \textit{associated Riemannian metric}
to the Lagrangian \eqref{Lag}. 

From the definitions (\ref{56}) we observe that $\beta=b_I y^I=y^0$, that
is, the linear one-form has only one non-vanishing coefficient $%
(b_0,b_1,b_2,b_3)=(1,0,0,0)$. Moreover, the determinant of the matrix $%
a_{IJ} $ can be explicitly written in terms of the initial data in the
Lagrangian \eqref{Lag}. 

For $a_{ij}=(1/2)m\delta _{ij}$, where $m$ is the mass of the
particle,  we find
\begin{eqnarray}
\det |a_{IJ}|&=&-\frac{1}{8}m^3\Bigg[\frac{e^2}{2mc^2}%
\left(A_1^2+A_2^2+A_3^2\right)+  \notag \\
&&\left(e\phi +V+V_Q\right)\Bigg].
\end{eqnarray}

For the sake of simplicity we will assume that the matrix $a_{IJ}$ is
positive defined, 
that is, we have a Riemannian metric on the manifold $\mathcal{M}=\mathbb{R}%
\times M$, called the {\it extendend configurations space},  with the local coordinates $(x^I)$. Under this assumption it
follows that the function $F$ in \eqref{F3} is a Kropina metric on $\mathcal{%
M}$, hence the theory presented in the previous Sections apply. Of course,
the case when the associated metric $a_{IJ}$ has a more general
signature, the theory can be studied in a similar manner.

In the absence of the external magnetic field, $A_i=0$, $i=1,2,3$, we obtain
for $a^{IJ}$, the inverse of the matrix $a_{IJ}$,  the simple form $a^{IJ}=-\left(e\phi+V+V_Q\right)^{-1}\times
\mathrm{diag}\left(1,a^{11},a^{22},a^{33}\right)$, where $%
a^{ii}=-(2/m)\left(e\phi+V+V_Q\right)$, $i=1,2,3$.

The $\alpha$-norm of $\beta$ is given by $
b^2=a^{IJ}b_I b_J=a^{00}=a_{11}a_{22}a_{33}/\det|a|>0$,
under the assumption that ${\det|a|}$ is positive. For later use, we remark
that $b^I:=a^{IJ}b_J=a^{I 0}b_0=a^{I 0}$.

We can summarize now our findings as the following

\textbf{Theorem 1.} \textit{The fundamental function $F=\frac{\alpha^2}{\beta}$%
, associated to the Lagrangian of the hydrodynamic representation of the
quantum mechanics, in the presence of external electromagnetic fields, is a
globally defined Kropina metric on the extended configurations space $%
\mathcal{M}$, where $\alpha^2=a_{IJ}y^Iy^J$ is the associated Riemannian
metric to the hydrodynamic Lagrangian of the quantum mechanics, and $%
\beta=y^0$. }

{\bf Remark.}
Kropina type Finsler metrics were associated  to a time-dependent Lagrangian in several studies, like, for example, \cite{Shib}, \cite{Rund2}, and \cite{AIM}. The starting point is the fundamental observation that {\it the Euler-Lagrange equations of the initial time-dependent Lagrangian actually coincide with the Euler-Lagrange equations of the associated Kropina metric} (see the references above for the proof of this result).

\subsection{The fundamental metric tensor of the Finslerian representation
of quantum mechanics}

Let us consider the fundamental tensor
\begin{equation}\label{69}
g_{IJ}=\frac{1}{2}\frac{\partial^2 F^2}{\partial y^I \partial y^J}=\frac{\partial F}{\partial y^I}\frac{\partial F}{\partial y^J}+F\frac{\partial ^2F}{\partial y^I\partial y^J},
\end{equation}
for the globally defined Kropina metric describing the geometric properties
of quantum hydrodynamics.

First we mention here an important result on the geometry of Kropina metrics
(\cite{YS1}), which states that the metric tensor of a Kropina space $F=%
\dfrac{\alpha^2}{\beta}$ is positive defined provided $\alpha$ is a positive
defined Riemannian metric \cite{YS1}.

Let us now turn back to the computation of the fundamental tensor of the Kropina metric %
\eqref{F3} associated to the Schr\"{o}dinger equation.
For the Finsler metric tensor we obtain the general expression
\bea
g_{IJ}&=&\left[B^2+2F\frac{T}{\left(y^0\right)^3}\right]\delta^0_I \delta^0_J+\left[\frac{B}{y^0}-\frac{F}{\left(y^0\right)^2}\right]\times \nonumber\\
&&\left(\delta ^0_IT_J+\delta ^0_JT_I\right)+2B\left(a_{0j}\delta ^0_I\delta ^j_J+a_{0i}\delta ^0_j\delta _I^i\right)+\nonumber\\
&&\left(2a_{0i}\delta ^i_I+\frac{T_I}{y^0}\right)\left(2a_{0j}\delta ^i_J+\frac{T_J}{y^0}\right)+F\frac{T_{IJ}}{y^0},
\eea
where we have denoted
\be
 B=a_{00}-\frac{T}{\left(y^0\right)^2}, T=a_{ij}(x)y^iy^j, T_{I}=\frac{\partial T}{\partial y^{I}},T_{IJ}= \frac{\partial ^2T}{\partial y^{I}\partial y^J}.
\ee


Explicitly, the Finslerian metric tensor components can be given as
\begin{equation}  \label{g_IJ1}
g_{00}  = a_{00}^2+4\frac{a_{0i}y^i}{(y^0)^3}T+3\frac{T^2}{(y^0)^4},
\ee
\be
g_{0i}  =2a_{0i}\left(a_{00}-\frac{T}{\left(y^0\right)^2}\right)-\frac{4a_{ij}y^j}{\left(y^0\right)^2}\left(\frac{T}{y^0}+a_{0j}y^j\right), \\
\end{equation}
\bea
\hspace{-0.3cm}g_{ij}&=&4a_{0i}a_{0j}+\frac{4\left(a_{ik}y^k\right)\left(a_{jl}y^l\right)}{\left(y^0\right)^2}+4a_{0j}\frac{a_{il}y^l}{y^0}+\nonumber\\
\hspace{-0.3cm}&&4a_{0i}\frac{a_{jl}y^l}{y^0}+2\left(\frac{T}{\left(y^0\right)^2}+\frac{2a_{0k}y^k}{y^0}+a_{00}\right)a_{ij}.
\eea

Alternatively, we can express the Finsler metric associated to the quantum hydrodynamical evolution as
\bea
g_{IJ}&=&2\left(\frac{\alpha}{\beta}\right)^2 a_{IJ}+\left(\frac{\alpha }{\beta}\right)^2\Bigg[4\alpha _I\alpha _J-\frac{3\alpha}{\beta } \times \nonumber\\ &&\left(\alpha _I b_J+\alpha _J b_I\right)+\frac{3\alpha ^2}{\beta ^2}b_Ib_J\Bigg].
\eea

A simple  computation shows that $\det |g_{IJ}|=24\left(\frac{\alpha}{\beta}\right)^8(1+d^2)\det|a_{IJ}|$,
where $d^2=\frac{3}{2}a^{IJ}\left(\alpha_I-\frac{\alpha}{\beta}b_I\right)\left(\alpha_J-\frac{\alpha}{\beta}b_J\right)$.
Hence the Kropina metric $g_{IJ}$ positive definiteness is equivalent to the positive definiteness of $a_{IJ}$.

\subsection{The local equations of the geodesics of the Kropina space}



We consider now the geodesics spray $G^I$ associated to the Kropina metric obtained in Eq.~\eqref{GJH_K}.
In this case $r_{IJ}=-\bar{\Gamma}_{IJ}^0,\quad s_{IJ}=0$,
where $\bar{\Gamma}_{IJ}^K$ are the Christoffel coefficients of the
associated Riemannian metric $\alpha$ on $\mathcal{M}$.


By components, we have
\begin{eqnarray}
G^0&=&\bar{G}^0 -\left(\frac{\beta}{b^2\alpha^2} {y^0} -\frac{1}{2b^2}
\right)r_{00}  \notag \\
&&= \bar{G}^0 +\left(\frac{\beta}{b^2\alpha^2} {y^0} -\frac{1}{2b^2} \right)%
\bar{\Gamma}_{IJ}^0y^I y^J,
\end{eqnarray}
\begin{equation}
G^i=\bar{G}^i -\left(\frac{\beta}{b^2\alpha^2} {y^i} \right)r_{00}=\bar{G}^i
+\left(\frac{\beta}{b^2\alpha^2} {y^i} \right)\bar{\Gamma}_{IJ}^0y^I y^J,
\end{equation}
where $\bar{G}^I=\dfrac{1}{2}\bar{\Gamma}^I_{JK}y^Iy^J$ are the spray
coefficients of the Riemannian metric $a_{IJ}$ obtained from the formulas %
\eqref{alpha G grI}, \eqref{alpha G grII}, and \eqref{alpha G grIII}, respectively.

Hence the Kropina metric geodesic equations are given by
\begin{equation}
\begin{split}
& \frac{d^2x^0}{d\tau^2}+2G^0\left(x(\tau),\frac{dx}{d\tau}\right)=0 \\
& \frac{d^2x^i}{d\tau^2}+2G^i\left(x(\tau),\frac{dx}{d\tau}\right)=0,
\end{split}%
\end{equation}
where $\tau$ is the $F$-unit parameter on the Kropina geodesics.

\subsection{The geodesics of the associated Riemann metric}

It is worth observing that the associated Riemannian metric
\eqref{associated
Riemann} is a metric on the extended configuration space that includes the
external potential, the electromagnetic potential,  and the
quantum potential generating the quantum effects during the evolution of the particle.  This represents {\it a new kind of Riemannian structure associated to such a Lagrangian function}.

The Levi-Civita connection coefficients of the
Riemannian metric $a_{IJ}$ are given by
\begin{equation}\label{Riemann Christo}
\bar{\Gamma}^I_{JK}=\frac{1}{2}a^{IL}\left( \frac{\partial a_{KL}}{\partial
x^J}+\frac{\partial a_{JL}}{\partial x^K}-\frac{\partial a_{JK}}{\partial x^L%
} \right).
\end{equation}

Generally, any second order tensor $\partial A^i/\partial x^j$ can be decomposed into a symmetric and anti-symmetric part,
\bea
\frac{\partial A_i}{\partial x^j}&=&\frac{1}{2}\left(\frac{\partial A_i}{\partial x^j}+\frac{\partial A_j}{\partial x^i}\right)+\frac{1}{2}\left(\frac{\partial A_i}{\partial x^j}-\frac{\partial A_j}{\partial x^i}\right)\nonumber\\
&&=\frac{1}{2}D_{ij}+\frac{1}{2}F_{ij},
\eea
where $D_{ij}$ represents the "strain" associated to the potential vector, while $F_{ij}$ can be interpreted physically as the strength of the magnetic field. The trace of $D_{ij}$ gives the divergence of the vector potential, $D_{ii}=\sum _{i=1}^3{\partial A_i/\partial x^i}$.

Thus for the Christoffel symbols of the associated Riemann metric we obtain
\begin{equation}\label{alpha G grI}
\bar{\Gamma}^0_{00} =\frac{1}{2}\left(a^{00}\frac{\partial a_{00}}{\partial
x^0}
-a^{0i}\frac{\partial a_{00}%
}{\partial x^i} \right),
\ee
\be
\bar{\Gamma}^i_{00}  =\frac{1}{2}\left(a^{i0}\frac{\partial a_{00}}{%
\partial x^0}
-a^{ij}\frac{%
\partial a_{00}}{\partial x^j} \right),
\end{equation}
\begin{equation}  \label{alpha G grII}
\bar{\Gamma}^0_{j0}
= \frac{1}{2} a^{00}\frac{\partial a_{00}}{\partial x^j}+a^{0i} \frac{e}{4c}F_{ij},
\ee
\be
\bar{\Gamma}^i_{j0} 
=\frac{1}{2} a^{i0}\frac{\partial a_{00}}{\partial x^j}+a^{ik} \frac{e}{4c}%
F_{kj},
\end{equation}
\begin{equation}  \label{alpha G grIII}
\bar{\Gamma}^0_{jk}
 =a^{00}\frac{e}{4c}D_{jk}-a^{00}A_i\gamma_{jk}^i,
 \ee
 \be
\bar{\Gamma}^i_{jk} 
=a^{00}\frac{e}{4c}D_{jk}+\gamma_{jk}^i,
\end{equation}
where we have denoted by $\gamma^i_{jk}$ the Levi-Civita connection coefficients of the Riemannian metric $a_{ij}$, $i,j\in\{1,2,3\}$. The geodesics of the Riemannian metric $a_{IJ}$ read now
\begin{equation}\label{92}
\frac{d^2x^0}{d\sigma^2} +\bar{\Gamma}^0_{IJ} \frac{dx^I}{d\sigma}\frac{dx^J%
}{d\sigma}=0,\;\;
\frac{d^2x^i}{d\sigma^2} +\bar{\Gamma}^i_{IJ} \frac{dx^I}{d\sigma}\frac{dx^J%
}{d\sigma}=0,
\end{equation}
where $\sigma$ is the $\alpha$-unit parameter, and $\bar{\Gamma}^K_{IJ}$ are
given above.

For a general Kropina metric $F=\dfrac{\alpha ^2}{\beta}$, it is easy to see that
$F$ is projectively equivalent to the Riemannian metric $\alpha$, i.e. the $F$-geodesics coincide with the $\alpha$-geodesics, if and only if $\beta$ is parallel with respect to the Levi-Civita connection of $\alpha$, that is $b_{I|J}=0$. In our case, the projectively equivalent condition is
$\bar{\Gamma}^0_{IJ}=0$, for all $I,J\in \{0,\dots,4\}$.

The geodesic equations (\ref{92}) describe the motion of a quantum particle in fields generated by an external (non-electromagnetic), electromagnetic, and quantum potential, in a purely "physical" geometry, generated by the potentials only. The derivative $d^2x^I/d\sigma ^2$ gives, from a physical point of view, the four-acceleration of the particle. Therefore the quantity $- \bar{\Gamma}^I_{JK}\frac{dx^J}{d\sigma} \frac{dx^K}{d\sigma}$ can be interpreted as the four-force acting on the particle as the result of the presence of the three types of potentials. Moreover, the tensor $a_{IJ}$ plays the role of the "potential" of the total field, since its derivatives determine the total field "intensity"  $\bar{\Gamma}^I_{JK}$.

\section{The Zermelo navigation problem in the hydrodynamics representation
of quantum mechanics}\label{sect5}

The navigation data $(h,W)$ on $\mathcal{M}$ corresponding to the Kropina
metric geometrizing quantum hydrodynamics are
\begin{equation}\label{h_alphabeta}
h_{IJ}=e^{\kappa(x)}a_{IJ},
W_I=\frac{1}{2}e^{\kappa(x)}b_I,
\end{equation}
respectively, where the conformal factor $e^{\kappa(x)}$ is obtained as
\begin{equation}  \label{k conformal fact}
e^{\kappa(x)}=\frac{4}{a^{00}}=\frac{4\det|a_{IJ}|}{a_{11}a_{22}a_{33}},
\end{equation}
where we assume that $a_{IJ}$ is positive defined.
More exactly, $W_I$ has only a nonzero component, $W_0=\frac{1}{2}e^{\kappa(x)}b_0=\frac{1}{2}e^{\kappa(x)}$
and $W_i=0$, for $i\in\{1,\dots,n\}$.

Taking into account that $h^{IJ}=e^{-k(x)}a^{IJ}$, we find
\begin{equation}\label{W^alpha}
W^I=h^{IJ}W_J=h^{I 0}W_0=\frac{1}{2}e^{\kappa(x)}h^{I 0}=\frac{1}{2}a^{I 0},
\end{equation}
and therefore the Kropina metric is determined by the Riemannian metric $%
h=h_{IJ}$ and the $h$-unit vector field $W=W^I\dfrac{\partial}{\partial x^I}$%
, $|W|_h=1$, defined on the manifold $\mathcal{M}$.

We point out that

\begin{enumerate}
\item - the Riemannian metric $h_{IJ}$ obtained in the navigation data above
is actually a conformal change of the extended Riemannian metric $a_{IJ}$.
Hence one can study the geometrical and physical properties of the
Riemannian metrics $a_{IJ}$ and $h_{IJ}$ by means of conformal
transformations and conformal geometry.

\item - the quantum wind $W$ is actually the vector field with the components
\begin{eqnarray}
&&W^I=\frac{1}{2\det|a_{IJ}|}\times \Bigg(a_{11}a_{22}a_{33}, -\frac{e}{2c}A_1 a_{22} a_{33}, \nonumber\\
&& -\frac{e}{2c}A_2a_{11}a_{33}
, -\frac{e}{2c}A_3a_{11}a_{22}\Bigg),
\end{eqnarray}
that is, the wind is given by  a combination of the mass field $a_{ij}$, and the
electromagnetic potential $A_i$.
\end{enumerate}

In the case of a single particle with constant mass, we have $a_{11}=a_{22}=a_{33}=m/2$, and the components of the quantum wind vector $W$ are given by
\be
W^I=\frac{\left(-1,\frac{e}{mc}\vec{A}\right)}{\left(e^2/mc^2\right)\left|\vec{A}\right|^2+2\left(e\phi+V+V_Q\right)},
\ee
where we have denoted by $\vec{A}=\left(A_1,A_2,A_3\right)$ the three dimensional vector potential of the magnetic field, and $\left|\vec{A}\right|^2=A_1^2+A_2^2+A_3^2$. The "timelike" component of the quantum wind is determined by $1/\det|a_{IJ}|$ only, while the spacelike components are proportional to the vector potential components, scaled by the determinant of the Riemannian metric. It is interesting to note that the wind is a function of the potentials only, and not on the fields, which are generally obtained as the gradients of the potentials. In the absence of the electromagnetic fields the quantum wind takes the form
\bea
W^I&=&\left(-\frac{1}{2\left(V+V_Q\right)},0,0,0\right)=\nonumber\\
&&\left(\frac{1}{2\left(-V+\frac{\hbar ^2}{2m}\frac{\Delta R}{R}\right)},0,0,0\right),
 \eea
where the real function $R$ is related to the wave function $\Psi \left(\vec{r},t\right)$ by the relation $\left|\Psi \left(\vec{r},t\right)\right|=R^2\left(\vec{r},t\right)$.  It is interesting to point out that the quantum wind is generated by the quantum potential $V_Q$, together with the external potential. When the magnitude of the quantum potential is much larger than that of the external potential, we obtain
\be
W^I\approx \frac{m}{\hbar ^2}\left(\frac{R}{\Delta R},0,0,0\right).
\ee

On the other hand if the vector potential of the magnetic field satisfies the condition $\left(e^2/mc^2\right)\left|\vec{A}\right|^2>>2\left(e\phi+V+V_Q\right)$, then the quantum wind is a purely classical quantity, which can be approximated as
\be
W^I\approx \left(-\frac{mc^2}{e^2\left|\vec{A}\right|^2},\frac{c}{e}\frac{\vec{A}}{\left|\vec{A}\right|^2}\right).
\ee

Hence the wind acting on the quantum particle is determined by the magnetic field potential only. The conformal factor $e^{\kappa(x)}$ is explicitly given by
\be
e^{\kappa(x)}=-4\left[\frac{e^2}{2mc^2}\left|\vec{A}\right|^2+\left(e\phi +V+V_Q\right)\right],
\ee
and it must satisfy the condition $e^{\kappa(x)}>0$. Such a conformal factor can always be obtained by taking into account that the vector potential of the magnetic field is not uniquely fixed, since if $\vec{A}$ is a vector potential, then $\vec{A}+\nabla \chi $ is also a vector potential, with $\chi $ an arbitrary function of the coordinates. Similarly, the electric field potential and the external potential can be rescaled. Therefore one can always choose a particular gauge for which the conformal factor is positive. For the components of the navigation metric we obtain the explicit components
\bea
h_{00}
\left(x^0,x^i\right)&=&4\left[\frac{e^2}{2mc^2}\left|\vec{A}\right|^2+\left(e\phi +V+V_Q\right)\right]\times \nonumber\\
&&\left(e\phi +V+V_Q\right),
\eea
\be
h_{0i}\left(x^0,x^i\right)=-2\frac{2}{c}A_i\left[\frac{e^2}{2mc^2}\left|\vec{A}\right|^2+\left(e\phi +V+V_Q\right)\right],
\ee
\be
h_{ij}\left(x^0,x^i\right)=-2m_{ij}\left[\frac{e^2}{2mc^2}\left|\vec{A}\right|^2+\left(e\phi +V+V_Q\right)\right].
\ee

Even if the components of the metric $a_{IJ}$ are functions of the spacelike coordinates only, or constants, the conformally transformed navigation metric is a function of both space and time coordinates. In the particular case of a neutral particle or in the absence of the electromagnetic interactions, $\phi \equiv 0$ and $\vec{A}\equiv 0$, and thus the conformal factor and the navigation metric take the simple forms
\be
e^{\kappa(x)}=-4\left(V+V_Q\right)=-4\left(V-\frac{\hbar ^2}{2m}\frac{\Delta R}{R}\right),
\ee
\be
h_{00}\left(x^0,x^i\right)=4\left(V+V_Q\right)^2=4\left(V-\frac{\hbar ^2}{2m}\frac{\Delta R}{R}\right)^2,
\ee
\be
h_{0i}\left(x^0,x^i\right)=0,i=1,2,3,
\ee
\bea
h_{ij}\left(x^0,x^i\right)&=&-2m\delta_{ij}\left(V+V_Q\right)=-2m\delta _{ij}\times \nonumber\\
&&\left(V-\frac{\hbar ^2}{2m}\frac{\Delta R}{R}\right), i,j=1,2,3.
\eea

The navigation metric is diagonal, and it is completely determined by the sum of the external potential, and of the quantum potential. Since $V+V_Q$ is the potential energy $E_p$ of the quantum hydrodynamic system, it turns out that all the components of the navigation metric are expressed in terms of the potential energy in the energy space. Similarly, the conformal factor $e^{\kappa(x)}$ is basically the potential energy of the quantum system. On the other hand, in the case of the metric $a_{IJ}$, only the component $a_{00}=-\left(V+V_Q\right)$ gives the potential energy of the quantum system, with the components $a_{ij}$ being proportional to the mass (or rest mass energy $mc^2$) of the system, assumed to be a constant. But all the components of the conformal metric $h_{IJ}$ are expressed in terms of the same physical variable, the potential energy of the quantum system.

Observe that the vector field $W$ has no zeros on $\mathcal{M}$. Indeed, if $%
W$ would have zeros, then in such a point, the functions $a^{I 0}$ must
vanish, for all $I\in \{0,1,\dots,n\}$, but this is not possible because
this would imply that in such a point $\det|a_{IJ}|= 0$. Also, at a zero point of $%
W$, the potential $e\phi +V+V_{Q}$ and $A_i$ must vanish simultaneously, and
this is not possible from physical point of view, since during the quantum evolution $V_Q\neq 0$.

Moreover, it easily follows that $a_{IJ}b^I b^J=a_{IJ}a^{I 0}a^{J 0}=a^{00}=b^2$, and therefore $h_{IJ}W^I W^J=1$. Hence our results are consistent with the general theory.


We can formulate now our findings by means of a {\bf Theorem} as follows.

{\bf Theorem 2.}
{\it The fundamental function
$F=\frac{\alpha^2}{\beta}:T({\rm R}\times M)\to {\rm R}$ given by \eqref{F3}, associated to the Lagrangian \eqref{Lag} describing the quantum hydrodynamical motion of a particle in the presence of external electromagnetic and non-electromagnetic fields, is the solution of the Zermelo's navigation problem on the Riemannian manifold $(\R\times M,h_{IJ})$ under the influence of the $h$-unit wind $W$.}

In other words, {\it
the time minimizing trajectories of a point on the space $\mathrm{R}\times M$
are not the Riemannian geodesics of $a_{IJ}$, but the geodesics of the
globally defined Kropina metric $(\mathrm{R}\times M,F)$.}

The Kropina metric obtained from the quantum Lagrangian Eq.~\eqref{Lag} is
therefore obtained by a rigid translation of the Riemannian metric $h_{IJ}$
 by the $h$-unit vector field $W$. Hence
the quantum Lagrangian is equivalent to a Kropina type
1-homogeneous Lagrangian, and this is obtained by the deformation of a
Riemannian conformal metric by means of $W$.

\section{Finslerian geometrization of the Schr\"{o}dinger equation for a
spinless uncharged particle}\label{sect6}

Let us consider the simple Lagrangian describing the evolution of a single
uncharged particle in the hydrodynamic representation of the Schr\"{o}dinger
equation, given by
\begin{equation}  \label{Lag1}
L\left(t,x,\dot{x}\right)=\frac{1}{2}\sum_{i=1}^3m({\dot{x}}^{i})^2-Q(t,x),
\end{equation}
where $Q(t,x)=V(t,x)+V_Q(t,x)$, with $V$ denoting the external potential,
while $V_Q$ is the quantum potential, responsible for the intrinsic quantum
features of the dynamic evolution.
The associated Kropina metric is given by $F=\frac{\alpha^2}{\beta}$, where $%
\alpha ^2=a_{IJ}y^Iy^J$ has the entries
\begin{equation}  \label{associated Riemann simple}
\begin{split}
a_{00}(x^0,x^i)& :=- {Q}, a_{0i} :=0, a_{ij} :=\frac{1}{2}m\delta_{ij}, i,j
\in \{1,2,3\}.
\end{split}%
\end{equation}
As for $\beta$, it
is given by $\beta=b_I y^I=y^0$, that is, the linear one-form has only one
non-vanishing coefficient $(b_0,b_1,b_2,b_3)=(1,0,0,0)$.

The determinant of $a_{IJ}$ is given by $\det |a_{IJ}|=-m^3Q/8$.
The associated Riemannian metric is positive
definite if and only if $Q<0$, that is we need to be in the setting where
the potential energy is positive. This can be always achieved by changing
the reference system.

We can conclude therefore our preliminary investigations of the geometric
description of the quantum hydrodynamics of a single particle by stating the
following

\textbf{Theorem 3.} \textit{The fundamental function $F=\frac{\alpha^2}{\beta}%
:T(\mathbb{R} \times M)\to \mathbb{R} $, associated to the Lagrangian %
\eqref{Lag1} describing the quantum motion of a neutral spinless particle, is a globally defined Kropina metric on the extended
configurations space $\mathcal{M}$, provided $Q<0$. 
 }

The inverse of the matrix $a_{IJ}$ is given by $a^{IJ}={\rm diag}\left(-1/Q,2/m,2/m,2/m\right)$.

\subsection{Geodesic evolution in the Finsler space}

The
Finsler metric function $F$ associated to the Lagrangian (\ref{Lag1}) is given by
\bea
&&F\left( x^{I},y^{I}\right) =\frac{\alpha ^{2}}{\beta }=\frac{ m\delta
_{ij}y^{i}y^{j}/2-Q\left(x^0,x^i\right) y^{0}  }{y^{0}}=\nonumber\\
&&\frac{
a_{00}\left(x^0,x^i\right)(y^{0})^{2}+T }{y^0}=a_{00}\left(x^0,x^i\right)y^0+\frac{T}{y^0},
\eea
where we have denoted $\alpha ^{2}=a_{00}(y^{0})^{2}+T$,$\quad
T=m\delta_{ij}y^{i}y^{j}/2$, and $\beta =y^{0}$, respectively.

Then, with the use of Eqs.~(\ref{69}), the fundamental tensor of the Kropina space describing the geometric
structure of the Schr\"{o}dinger equation for a spinless particle is obtained as
\begin{eqnarray}\label{104}
g_{IJ} &=&\left[ a_{00}^{2}\left( x^{I}\right) +\frac{3T^{2}}{\left(
y^{0}\right) ^{4}}\right] \delta _{I}^{0}\delta _{J}^{0}-\frac{2T}{\left(
y^{0}\right) ^{3}}\left( T_{I}\delta _{J}^{0}+T_{J}\delta _{I}^{0}\right) +\nonumber\\
&&\frac{1}{\left( y^{0}\right) ^{2}}T_{I}T_{J}+
T_{IJ}\left[ a_{00}\left(
x^{I}\right) +\frac{T}{\left( y^{0}\right) ^{2}}\right] .
\end{eqnarray}%

Explicitly, the components of the Finslerian metric tensor are given by
\begin{equation}
g_{00}=Q^{2}+3\frac{T^{2}}{(y^{0})^{4}}, \quad
g_{0i} =-2m\frac{\delta _{ij}y^j}{\left(y^0\right)^3}T ,
\ee
\be
g_{ij} =m\delta _{ij}\left( a_{00}+\frac{T}{(y^{0})^{2}}\right) +m^2\frac{%
\delta_{il}y^{l}\delta_{jk}y^{k}}{(y^{0})^{2}}.
\label{g_IJ}
\end{equation}

The spray coefficients  describing the evolution of a neutral spinless quantum particle in the associated Kropina space are defined as
\begin{equation}
G^{I}\left( x,y\right) =\frac{1}{2}\Gamma _{JK}^{I}y^{J}y^{K}=\frac{1}{%
4}g^{IJ}\left( \frac{\partial ^{2}F^{2}}{\partial x^{K}\partial y^{J}}y^{K}-%
\frac{\partial F^{2}}{\partial x^{J}}\right) ,
\end{equation}%
where $\Gamma _{JK}^{I}$ are the analogues of the Christoffel symbols of the
Riemann geometry.
By taking into account the explicit form of the Finslerian metric tensor coefficients as given by Eqs.~(\ref{104}), we  obtain
\begin{eqnarray}
\Gamma _{JK}^{I}\left( x,y\right) &=&g^{IL} a_{00} \Bigg\{%
\frac{\partial }{\partial x^{K}}\left[ a_{00} \delta
_{L}^{0}\delta _{J}^{0}+\ln \sqrt{\left|a_{00}\right| }T_{JL}\right] +
\notag \\
&&\frac{\partial }{\partial x^{J}}\left[ a_{00} \delta
_{L}^{0}\delta _{K}^{0}+\ln \sqrt{\left|a_{00}\right| }T_{LK}\right] -
\notag \\
&&\frac{\partial }{\partial x^{L}}\left[ \left|a_{00}\right| \delta
_{J}^{0}\delta _{K}^{0}+\ln \sqrt{\left|a_{00}\right|}T_{JK}\right] %
\Bigg\}.
\end{eqnarray}%
By denoting
\begin{equation}
\tilde{g}_{IJ}=a_{00}\left( X^{I}\right) \delta _{I}^{0}\delta _{J}^{0}+\ln
\sqrt{a_{00}\left( X^{I}\right) }T_{IJ},
\end{equation}%
we find for the spray coefficients the expressions
\begin{equation}
G^{I}(x,y)=\frac{1}{2}g^{IL} a_{00} \left[
\frac{\partial \tilde{g}_{LJ}}{\partial x^{K}}+\frac{\partial \tilde{g}_{LK}%
}{\partial x^{J}}-\frac{\partial \tilde{g}_{JK}}{\partial x^{L}}\right]
y^{J}y^{K},
\end{equation}
which allows us to write the equations of the geodesics of the Finslerian geometric formulation of the quantum hydrodynamics as
\be
\frac{d^2x^I}{d\lambda ^2}+2G^{I}(x,y)=0.
\ee

\subsection{Geodesic equations in the associated Riemann space}

Let us observe that the Levi-Civita connection coefficients of the
Riemannian metric $a_{IJ}$ given by \eqref{Riemann Christo} can be written as

\bea\label{alpha G grI2}
\bar{\Gamma}_{00}^{0}&=&\frac{1}{2}a^{00}\frac{\partial a_{00}}{\partial x^{0}}%
,\bar{\Gamma}_{00}^{i}=-\frac{1}{2}a^{ij}\frac{\partial a_{00}}{\partial
x^{j}}, \bar{\Gamma}_{j0}^{0}=\frac{1}{2}a^{00}\frac{\partial a_{00}}{\partial x^{j}}%
, \nonumber\\
\bar{\Gamma}_{j0}^{i}&=&0,  \label{alpha G grII2}
\bar{\Gamma}_{jk}^{0}=0,\bar{\Gamma}_{jk}^{i}=\gamma _{jk}^{i}=0,
\eea
where we have denoted by $\gamma^i_{jk}$ the
Levi-Civita connection coefficients of the Riemannian metric $a_{ij}$ (obviously $\gamma^i_{jk}=0$ because $%
a_{ij}$ are constants).

The geodesics of the Riemannian metric $a_{IJ}$ are given by
\begin{equation}
\frac{d^{2}x^{0}}{d\sigma ^{2}}+\frac{1}{2Q}\left( \frac{\partial Q}{%
\partial x^{0}}\frac{dx^{0}}{d\sigma }+2\frac{\partial Q}{\partial x^{i}}%
\frac{dx^{i}}{d\sigma }\right) \frac{dx^{0}}{d\sigma }=0,  \label{96}
\end{equation}%
\begin{equation}
\frac{d^{2}x^{i}}{d\sigma ^{2}}+\frac{1}{m}\frac{\partial Q}{\partial x^{j}}%
\delta ^{ij}\left( \frac{dx^{0}}{d\sigma }\right) ^{2}=0,  \label{97}
\end{equation}%
where $\sigma $ is the $\alpha $-unit parameter. 
Eq.~(\ref{96}) can be rewritten as
\be
\frac{d}{d\sigma}\ln \left(\frac{dx^0}{d\sigma}\sqrt{Q}\right)=-\frac{1}{2}\left.\left(\frac{1}{Q}\frac{\partial Q}{\partial x^i}\right)\right|_{\left(x^0(\sigma),x^i(\sigma)\right)}\frac{dx^i}{d\sigma}.
\ee
By integrating the above equation from zero to sigma 
and by substituting the resulting expression of $dx^0/d\sigma$  in Eq.~(\ref{97}), it follows that the equation of evolution of the space-like coordinates $x^i$ of the spinless neutral quantum particle in the associated Riemann space is described by the equation
\bea
\hspace{-0.8cm}&&\frac{d^{2}x^{i}}{d\sigma ^{2}}+\frac{C^2}{m}\frac{\partial Q}{\partial x^{j}}%
\delta ^{ij}\times \nonumber\\
\hspace{-0.8cm}&&\exp\left[-\int_0^{\sigma}{\left.\left(\frac{1}{Q}\frac{\partial Q}{\partial x^i}\right)\right|_{\left(x^0(\sigma),x^i(\sigma)\right)}\frac{dx^i}{d\sigma}d\sigma}\right]=0,
\eea
where $C$ is an arbitrary integration constant. Once the explicit form of the potential $Q$ is known, the solutions of the geodesic equations in the associated Riemann space can be obtained by using either analytical or numerical methods.

\subsection{The Zermelo navigation problem}

We also observe that we can formulate the Zermelo's navigation problem for
this simple Lagrangian. Indeed, for the navigation data $(H,W)$, where $h_{IJ} =e^{k(x)}a_{IJ}=-4Q{\rm diag}\left(-Q,m/2,m/2,m/2\right)$ and $W_I  = (W_0,0,0,0)=(-Q/2,0,0,0)$
we have the following {theorem}. Here we have used $e^{k(x)}=4/a^{00}%
=-4Q$.

\textbf{Theorem 4.}\textit{\ The fundamental function $F=\frac{\alpha^2}{\beta}:T(%
\mathbb{R} \times M)\to \mathbb{R} $ associated to the Lagrangian %
\eqref{Lag1}, is the solution of the Zermelo's navigation problem on the
Riemannian manifold $(\mathbb{R} \times M,h_{IJ})$ under the influence of
the $h$-unit wind $W$. In other words, the time minimizing trajectories of a
point on the space $\mathbb{R} \times M$ are not the Riemannian geodesics of
$a_{IJ}$, but the geodesics of the globally defined Kropina metric $(\mathbb{%
R} \times M,F)$. }

For the Christoffel symbols in the $h$-space we obtain
\begin{equation}
\overset{(h)}{\Gamma}\ ^0_{00}  =\frac{5}{2}\frac{1}{Q}\frac{\partial Q}{%
\partial t},\;
\overset{(h)}{\Gamma}\ ^0_{ij}  =0,\;
\overset{(h)}{\Gamma}\ ^0_{0j}  =\frac{3}{2}\frac{1}{Q}\frac{\partial Q}{%
\partial x^j}.
\end{equation}
Then the Killing conditions
read
\begin{equation}
\begin{split}
 \frac{\partial Q}{\partial t}=0,
 \frac{\partial Q}{\partial x^j}=0,
\end{split}%
\end{equation}
that is, the wind $W$ is Killing for the globally defined Kropina
metric induced by the Lagrangian \eqref{Lag1} if and only if $Q$ is a constant. For the sake of convenience we will consider this constant to be negative $Q<0$.

In this case we can apply the general theory presented in \cite{SSY} to this Kropina metric, however the following peculiarities appear:
\begin{enumerate}
    \item - since the potential $Q$ is constant, the Lagrangian  \eqref{Lag1} is a time independent Lagrangian, so the new coordinates $(x^0,y^0)$ are not anymore geometrically intrinsic;
    \item - since all the entries of the matrix $a_{IJ}$ are constants, it follows that the Riemannian metric $\alpha$ is Euclidean, the Levi-Civita connection  coefficients $\bar{\Gamma}^K_{IJ}$ all vanish, and the 1-form $\beta$ is parallel. This means that the $F$-geodesics coincide with the $\alpha$-geodesics. In other words, the $F$-geodesics are straight lines.
\end{enumerate}
Even though the Riemannian metric and the unit Killing vector field $W$ are both complete, it doesn't mean that we can join any two points by an $F$-geodesic.

\section{Discussions and final remarks}\label{sect7}

In the present paper we have considered an alternative geometric perspective of one of the fundamental fields of theoretical physics, quantum mechanics. Even that quantum mechanics is intrinsically a geometric theory, formulated in a Hilbert space, its structure is completely different from the space-time geometrical approach that was so successful in the description of gravitational interaction. The geometrizations of quantum mechanics in its standard formulation, even if they use some of the theoretical tools of Riemannian geometry, still present fundamental differences as compared to the space-time geometry of special and general relativity. In our work we have adopted a different approach to the geometrization of quantum mechanics, namely, we have adopted as a starting point the so-called hydrodynamical (or Madelung) formulation of quantum mechanics. In this formulation, extensively used in many fields of physics, the quantum motion of the particle is described by the standard equations of classical fluid dynamics (continuity and Euler equations), in the presence of a quantum potential, which essentially determines the quantum properties of the motion. The hydrodynamical interpretation is a standard theoretical tool in the physics of quantum gases, superfluidity, and Bose-Einstein Condensation theory, and it allows the realization of a deeper connection between experiment and theory.  This formulation is also at the basis of the de Broglie-Bohm pilot wave (deterministic) interpretation of quantum mechanics (see \cite{Holland} for an in depth analysis of the causal interpretation of the quantum theory). Hence in the hydrodynamic interpretation of the Schr\"{o}dinger equation the evolution of the quantum particle is described by the motion of a classical fluid in the presence of a specific potential. However, {\it the Madelung representation of the wave mechanics can be seen as a mathematical tool, which does not change the physical interpretation of the quantum mechanics.}

On the other hand any space-time type geometrization of quantum mechanics requires the introduction of the concept of quantum trajectory \cite{Wyatt}. In the Bohmian interpretation of quantum mechanics the quantum trajectory plays the role of a hidden variable,  with the path being the fundamental dynamical
variable of the theory \cite{Bohm1, Bohm2}, while the wave function evolves in time along the Lagrangian motion of the path \cite{Foskett}. However, it is important to point out that the particle dynamics along the Bohmian trajectories is not equivalent to the point particle evolution along classical Newtonian trajectories \cite{Foskett}. One reason is that in principle the classical behavior should be obtained by adopting a point-like initial density of the Dirac delta function type for the density,  and then integrating the Euler-Lagrange equation over a reference density. However, the structure of the quantum potential does not allow this type of initial condition \cite{Foskett}.

Once the definition of the quantum trajectory is introduced, the equations of motion of the quantum particle can be derived from an action principle, which, in the presence of electromagnetic fields, contains the standard classical terms plus the quantum potential, responsible for the nonclassical features of the motion. The Lagrangian function can be taken as the starting point for the geometrization of quantum hydrodynamics. After performing the homogenization procedure of the Lagrangian introduced in \cite{Shib}, it turns out that the Finsler function associated to the Lagrangian of the quantum mechanics takes the form of the Kropina function \cite{Krop1,Krop2}, which is a particular form of the general $(\alpha, \beta)$ metrics \cite{Mat1, Mat2}, whose geometric properties have been extensively studied in the mathematical literature. The Kropina metric can also be expressed in terms of a Riemann metric $a_{IJ}$ associated to the quantum Lagrangian, and which is entirely determined by the components of the external potentials (including the quantum potential and the electromagnetic potentials). Once the Finsler metric function is known, the Finslerian metric tensor components can be obtained in a straightforward way, as well as the geodesic equations, which are equivalent to the quantum equations of motion of the particle. It is interesting that similar equations of motion, fully determined by the external potentials, can be also written down in the associate Riemann space, with metric $a_{IJ}$. They describe the motion of the quantum particle in the presence of a particular force generated via the Christoffel symbols associated to the Riemann metric.

The geometrization of the hydrodynamic formulation of the quantum mechanics also opens some new perspectives on the quantum Zermelo navigation problem, a subject of major interest in quantum computation. Once the geometric Finsler type structure of the quantum hydrodynamic flow is known, the Zermelo navigation data can be easily constructed by means of a conformal transformation of the associated Riemannian metric $a_{IJ}$, and of the vector $b_I$, respectively, The quantum wind can be obtained  as a function of the external and quantum potential, respectively. The Kropina fundamental function of quantum hydrodynamics is a solution of the Zermelo problem, a result that may find some useful applications  in the study of quantum speed limits and quantum information transfer.

One of the main advantages of the Finslerian geometrization of quantum mechanics is that the entire formalism is constructed in the ordinary space-time geometry, with all geometrical quantities functions of the external physical variables (field potentials). Even that the expressions of the Finslerian metric tensors and of the geodesic sprays are relatively complicated, they can be handled from a mathematical point of view in a much easier way than their purely quantum counter parts defined on a Hilbert space. Hence the results obtained in the present paper may provide some deeper insights into the complex relation between the classical and quantum worlds, as well as on the geometrical structures underlying the quantum mechanical dynamics.
\section*{Acknowledgments}

S. V. S. and T. H. would like to thank the School of Mathematics and the Yat Sen School of the Sun Yat Sen University
in Guangzhou, P. R. China, for the kind hospitality offered during the early stages of the preparation of this work.






\end{document}